\documentclass[letterpaper, 10 pt, conference]{ieeeconf}  % Comment this line out
\IEEEoverridecommandlockouts                              % This command is only
\usepackage{cite}
\usepackage{amsmath,amssymb,amsfonts}
\usepackage{algorithmic}
\usepackage{graphicx}
\usepackage{textcomp}
\usepackage{xcolor}
\usepackage{comment}
\usepackage{amsmath}
\usepackage{mathtools}
\usepackage{subcaption}
\usepackage{multirow}
\usepackage{siunitx}
\usepackage[hyphens]{url} % Allows line breaks in long URLs
\usepackage[colorlinks, allcolors=black]{hyperref} % Makes links clickable and blue
\def\BibTeX{{\rm B\kern-.05em{\sc i\kern-.025em b}\kern-.08em
    T\kern-.1667em\lower.7ex\hbox{E}\kern-.125emX}}

\usepackage[symbol]{footmisc}

\begin{document}

\title{\LARGE \bf A Closed-Loop Model of an Anion Exchange Membrane Electrolyser Based on Operational Data}

\begin{comment}
\author{
\IEEEauthorblockN{Maiken Omtveit}
\IEEEauthorblockA{\textit{Department of Electrical Power Engineering} \\
\textit{Norwegian University of Science and Technology (NTNU)} \\
\textit{ABB AS} \\
Trondheim, Norway \\
maiken.omtveit@ntnu.no}
\and
\IEEEauthorblockN{Lisa Marie Dannappel}
\IEEEauthorblockA{\textit{Department of Wind and Energy Systems} \\
\textit{Technical University of Denmark (DTU)} \\
\textit{ABB AS} \\
Trondheim, Norway \\
lisadan@dtu.dk}
\and
\IEEEauthorblockN{Kjetil Uhlen}
\IEEEauthorblockA{\textit{Department of Electrical Power Engineering} \\
\textit{Norwegian University of Science and Technology (NTNU)} \\
Trondheim, Norway \\
kjetil.uhlen@ntnu.no}
\and 
\IEEEauthorblockN{Shi You}
\IEEEauthorblockA{\textit{Department of Electrical Power Engineering} \\
\textit{Department of Wind and Energy Systems} \\
\textit{Technical University of Denmark (DTU)} \\
Trondheim, Norway \\
shyo@dtu.dk}
}
\end{comment}

\author{ Maiken Borud Omtveit$^{\dagger,\ddagger}$, Lisa Marie Dannappel$^{\dagger\dagger}$, Shi You$^{\dagger\dagger}$, Kjetil Uhlen$^{\ddagger}$% 
\thanks{$^{\dagger}$R\&D Automation and Electrification, ABB AS, Fornebu, Norway. {\tt\small maiken.omtveit@no.abb.com}}% 
\thanks{$^{\ddagger}$Department of Electric Energy, Norwegian University of Science and Technology, Trondheim, Norway.}% 
\thanks{$^{\dagger\dagger}$Department of Wind and Energy Systems, Technical University of Denmark, Lyngby, Denmark.} 
\thanks{This manuscript has been accepted for publication in the Proceedings of IEEE ISGT 2026.}
}

\maketitle

\begin{abstract}
The main contribution of this work is to identify model parameters for an Anion Exchange Membrane (AEM) system directly from measured operational data, enabling their use in power system studies. Due to the relatively low technology readiness level (TRL) of AEM electrolysers, only limited literature reports operational parameters supported by openly available experimental data. By capturing the realistic dynamic behaviour and power consumption of the AEM electrolyser, this model allows for more accurate evaluation of system topologies and control strategies for green hydrogen production.

\begin{comment}
This paper aims to describe the behaviour of an AEM using data from a particular AEM electrolyser setup installed at the Power-to-X Laboratory at the Technical University of Denmark (DTU). Operational data was collected over one month in fall 2025, and the data was processed to find the polarisation characteristic, a model of the current dynamic and a lumped model for the auxiliary equipment. The model was validated on operational data. 
\end{comment}

This paper presents a model of an AEM electrolyser based on experimental data from Enapter modules installed in the Power-to-X (PtX) Laboratory at the Technical University of Denmark \cite{powerlab_ptx}. The model comprises a polarisation curve and a model of current dynamics during operation. %, and a lumped representation of the balance of plant. 
It is validated against operational data and can be applied in grid-balancing studies as well as in assessments of how electrolyser dynamics influence the power system.

The results suggest that the AEM technology is suitable for services such as smoothing wind farm power output and provide system services to support frequency balancing. The findings highlight the importance of considering the constraints of the process when designing for flexible load operation, and consideration of the interaction between the power control and process control systems. 

\end{abstract}


% \begin{highlights}
% \item Power-to-hydrogen can help decarbonize heavy industry and shipping sectors
% \item Study compares power-to-hydrogen systems in Denmark, the U.S., and China
% \item Regional factors impact cost, efficiency, and performance of hydrogen systems
% \item Customized designs are key to improving hydrogen system cost-effectiveness
% \item Focus on system scale, design choices, configuration, and operational strategies
% \end{highlights}

% Keywords
%Keywords: Anion Exchange Membrane Electrolyser \sep Modelling  \sep Hydrogen \sep Power-to-X 

\maketitle

% Main text

% \section*{Nomenclature}
% \subsection*{Abbreviations}
% \begin{tabbing}
%     PtX \hspace{2cm} \= Power-to-X \\
%     H$_2$ \> Hydrogen \\
% \end{tabbing}

% \subsection*{Symbols}
% \begin{tabbing}
%     OPEX$^{fix}_t$ \hspace{2cm} \= Annual fixed operations and maintenance costs in year $t$ \\
%     OPEX$^{var}_t$ \> Annual variable operations and maintenance costs in year $t$ \\
%     $t$ \> Year \\
% \end{tabbing}

% \subsection*{Symbols}
% \begin{tabbing}
%     $t$ \hspace{2cm} \= Year \\
%     OPEX$^{fix}_t$ \> Annual fixed operations and \\
% \end{tabbing}

\section{Introduction}\label{sec:intro}
Although AEM technology currently has low TRL \cite{IEA_AEM_TRL_2024}, it is a promising technology for green hydrogen production. Its ramp rate is comparable to other fast-ramping electrolyser technologies such as Proton Exchange Membrane (PEM) electrolysers, whereas the cost is significantly lower as it does not rely on expensive stable materials to endure the typical corrosive environment of protons like in PEM electrolysers \cite{GAO2025100563}. Hence, its characteristics make it a good candidate for integration with intermittent renewable energy. Models of alkaline and PEM electrolysers are quite well established, however there are currently few dynamic models available for AEM technology. The contribution of this paper is to establish a closed-loop AEM model of the current dynamics and the polarisation curve. The model is closed-loop in the sense that the process and converter control is protected by IP and viewed as a black box.
%in the low-margin business that Power-to-X is. 

%%%%%%%%%%%%%%%%%%%%%%%%%%%%%%%%%%%%%%%%%%%%%%%%%%%%%%%%%%%%%%%%%%%%%%%%%%%%%%%%%%%%%%%%%%
\label{sec:background}
Electrolyser modelling for power system studies encompasses a wide range of models that fit different purposes. The behaviour of different types of electrolyser technologies (alkaline, proton exchange membrane, solid oxide and AEM) resemble one another, so models in literature can, depending on the level of detailed, be generalized and tuned to fit each technology and vendor. The models differ with respect to modelling framework and time scale, which is usually dictated by the problem one wants to solve. 

There have been techno-economic studies on operation of AEMs that have been solved using an optimization problem formulation representing the process dynamics as simplified constraints to the objective function \cite{GUL2023117025, GULAY2025119797}. These models are typically used when the problem revolves around planning the production, and uses large time steps where average values are more important than fast dynamics. Dynamics are commonly disregarded or simplified as constant ramp rates.
% Including varying details in the modelled constraints affects the result of the optimization. As one study demonstrated, including more details can improve the final cost \cite{andreamodel}. 

A second approach is to use first principle modelling, where the dynamics are based on physical laws. This provides higher temporal resolution and physical insights into the system behaviour, and is useful for example when studying the power system dynamics and control. For an electrolyser system, this can be done with co-simulation of a process simulator and a power system simulator, to capture the dynamics of both the process and the power systems. It is ultimately the controller which decides the dynamics of operation, and considerations must be done with respect to aspects such as performance, degradation, efficiency etc. 

A third approach is to use experimental data of an existing system to directly identify a model of the electrolyser. In this case, both the model structure and parameters are dependent on the data. This is typically done using machine learning algorithms, and the models can be useful when implementing control both on high level planning level and low level fast control, depending on the model.

Often several approaches to modelling are combined, as in \cite{residuals} which employed a hybrid approach by first identifying an equivalent double branch RC-circuit then training a shallow neural network on residual errors to improve the model. Following E. P. Box's well-known principle, all models are inaccurate, but some can still be useful. In this paper, the investigated research question is how to model an AEM electrolyser's voltage and current for second-level time step simulation, based on experimental data. The motivation is to have a realistic representation of AEM electrolyser dynamics to be used in power system simulations.

\begin{comment}
As AEMs have similarities to other electrolysers, some electrolyser models can be applied to AEM technology. 
The method developed by Ulleberg for alkaline electrolysers has 
- General electrolyser models that can be applied to AEM and what they contain.
- Specific AEM electrolyser models (make graph of existing literature on this to prove contribution of paper?). 
- Shortcomings of different models, and purpose of different detailed levels. 
- 5-10 papers on modelling.
- polarisation curve, current dynamics, hydrogen production, thermodynamic model, degradation
- Introduce Ulleberg method.
- Refer to paper by Pierluigi Mancarella, and Nikita.
- Faradays law
\end{comment}
\vspace{-1pt}

%%%%%%%%%%%%%%%%%%%%%%%%%%%%%%%%%%%%%%%%%%%%%%%%%%%%%%%%%%%%%%%%%%%%%%%%%%%%%%%%%%%%%%%%%%
\section{Methodology}\label{sec:method}
\vspace{-1pt}
\subsection{Electrolyser Setup Description}\label{sec:setup}
\begin{figure}
\centerline{\includegraphics[width=0.3\textwidth]{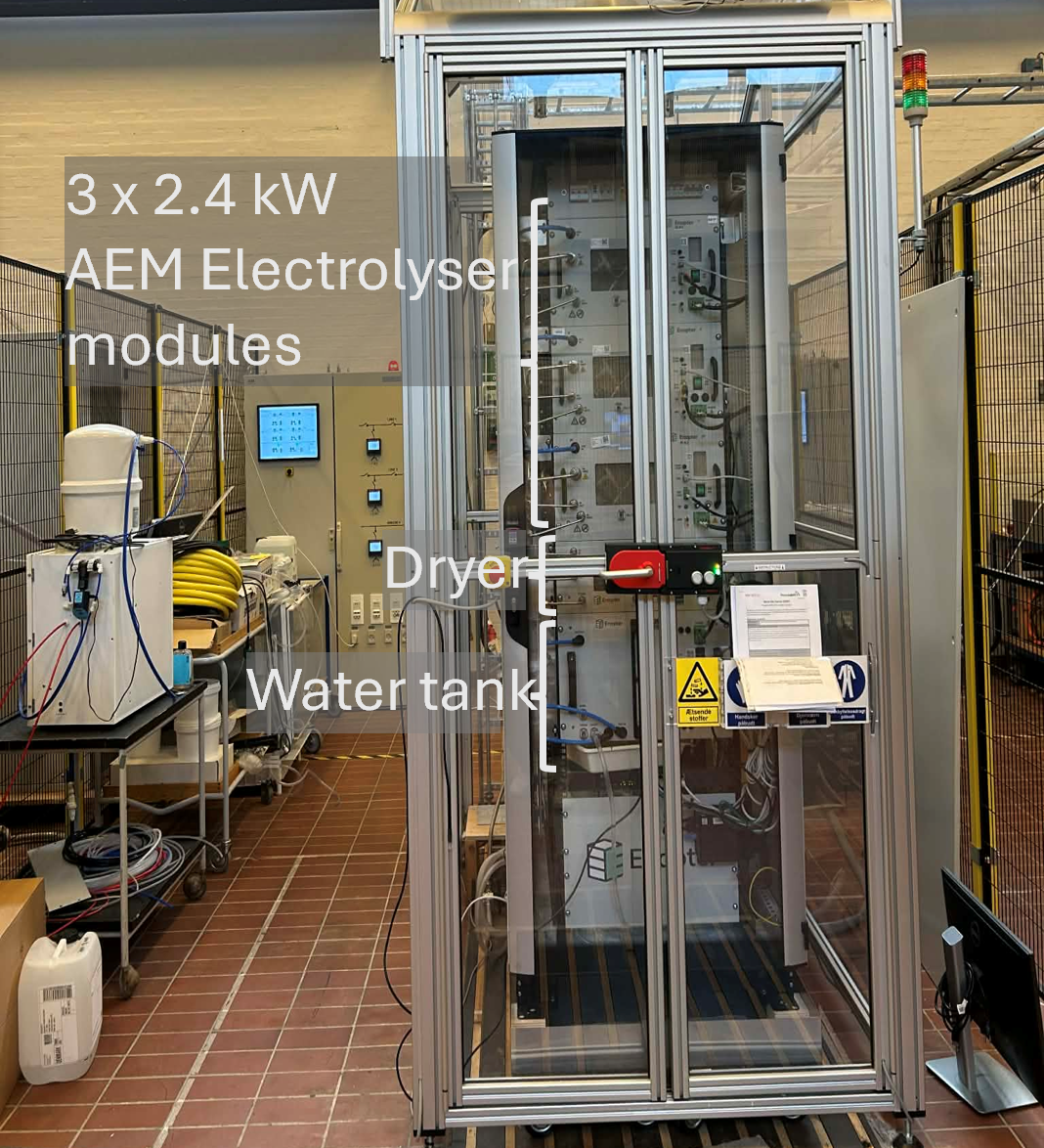}}
\caption{\label{fig:setup}Electrolyser Setup in DTU PtX Laboratory on the right, on the left a picture of the inside of the electrolyser module.}
\vspace{-18pt}
\end{figure}

The electrolyser setup, as shown in Fig. \ref{fig:setup}, is installed in the PtX Laboratory at the DTU Lyngby campus. It consists of three Enapter AEM electrolyser modules, with connected auxiliary process equipment, safety equipment and measurement systems. The setup has been described and used in previous research \cite{Anastasiya,Marius,Stoyan}.

The AEM electrolyser modules are all the 2.4 kW Enapter 4.1 model with air cooled AC \cite{EnapterHandbook2025}. It uses potassium hydroxide (KOH) electrolyte and there is an internal rectifier in each module. Inside each module there is a stack consisting of 23 cells. The auxiliary equipment for the Enapter modules consists of a central water tank module \cite{EnapterWaterTank2025} and a dryer \cite{EnapterDryer2025}. 
% Maybe include drawing of how it is connected
The external water supply and power supply are installed outside the cabinet. The water supply is deionized water that is manually filled into a tank and pumped into the central water tank whenever the pressure difference downstream of the tank reaches a threshold. There are two three-phase power supplies from the local grid, and each phase is connected to a part of the electrolyser system. On the first power supply each phase is connected to one of the three electrolyser modules. On the second power supply, one phase is for the dryer, one is for the water tank, and one is for auxiliary equipment such as screens.

% \subsubsection{Safety System}
There is one safety system built into the Enapter modules, and another installed by lab technicians at DTU. The Enapter modules are installed inside a safety cabinet with ventilation. Likewise, there is one sensor system built into the Enapter modules on the DC side with a measurement reading every 1 second, and one installed using DTUs sensors on the AC side. The sensors inside the Enapter modules cannot be accessed from the outside. 
% \subsection{Data Collection}\label{sec:data}

% \subsubsection{Uncertainties}
%Uncertainties in the setup include measurement noise and uncertainties in the sensors. 
\vspace{-1pt}

\subsection{Procedure of Data Collection}\label{Procedure}
The procedure for data collection was to run pre-defined scenarios of load setpoint changes on each of the electrolyser modules. The scenarios involved setpoint changes of varying step size and varying settling time. The three electrolyser modules were run both separately and together. The test runs and data collection was conducted 11 days during three weeks in October 2025, of which 10 are used for developing the model, and one is used for model validation. From the data, polarisation curves and step responses were extracted. 

% Make this figures prettier and more similar, maybe all in one figure even
\begin{comment}
\begin{figure}
\centering
\begin{subfigure}[b]{0.5\textwidth}
  \includegraphics[width=1\linewidth]{Images/newplot.png}
  \caption{}
  \label{fig:Ng1} 
\end{subfigure}
\medskip % insert a bit of vertical whitespace
\begin{subfigure}[b]{0.5\textwidth}
  \includegraphics[width=1\linewidth]{Images/TimeSeries.png}
  \caption{}
  \label{fig:Ng2}
\end{subfigure}
\caption[Setpoint schedule and corresponding measured current and voltage]{%
(a) Setpoint schedule for electrolyser in percentage.
(b) Time series of current and voltage of the electrolyser corresponding to the setpoint schedule.}
\label{fig:setpoints}
\end{figure}
\end{comment}

\begin{comment}
\begin{figure}
\centerline{\includegraphics[width=0.5\textwidth]{Images/oct22_setpoint_current_voltage_scientific_steady_overlay.png}}
\caption{\label{fig:setpoints}Example of setpoint schedule for electrolyser in percentage and measured current and voltage of the electrolyser stack.}
\vspace{-9pt}
\end{figure}
\end{comment}
\vspace{-1pt}

%%%%%%%%%%%%%%%%%%%%%%%%%%%%%%%%%%%%%%%%%%%%%%%%%%%%%%%%%%%%%%%%%%%%%%%%%%%%%%%%%%%%%%%%%%
\section{Data Analysis and Parametrization}\label{sec:analysis}

% \subsection{Typical Model Parameters}
\begin{figure}
\centerline{\includegraphics[width=0.45\textwidth]{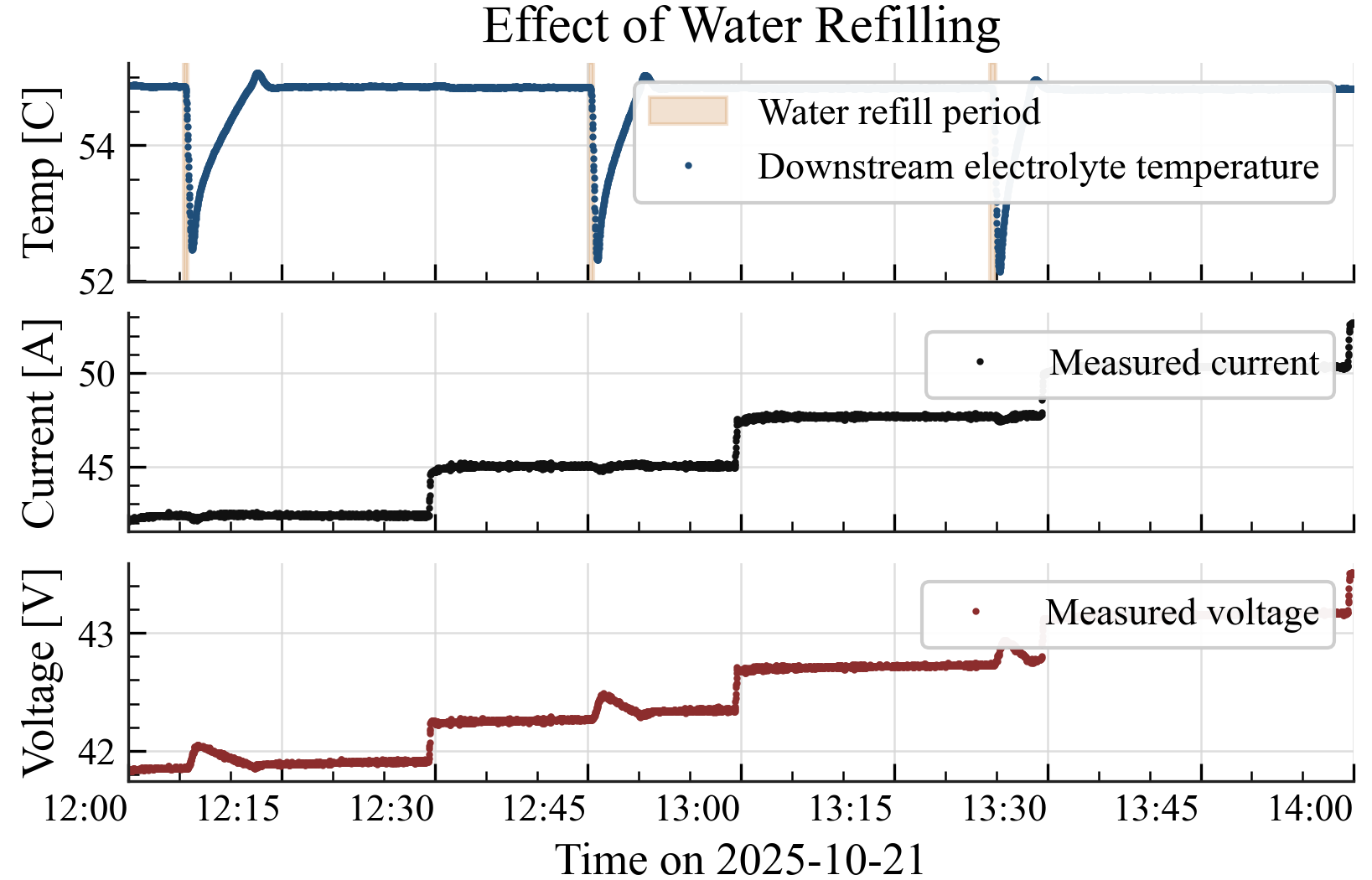}}
\caption{\label{fig:dips}Electrolyte temperature and stack current and voltage of electrolyser. The small dips in temperature correlate with the small spikes in voltage.}
\vspace{-18pt}
\end{figure}

\subsection{Modelling Method}
The polarisation curve lies in the chemical domain, whereas the control of the current and voltage is in the electrical domain. The polarisation curve is commonly modelled as a sum of three overpotentials - reaction kinetics, ohmic losses and mass transfer resistance, a detailed process model of which has been developed \cite{AEM_num, YAN2026153793}. A study on the same Enapter module as investigated in this paper finds that it exhibits polarisation behaviour close to theoretical models \cite{UiS}.

The parameters for the electrolyser model found in this study are given in Table \ref{tab:params}. The parameters that affect the power consumption of the electrolyser during operation is the current and the voltage. Hence, the approach to modelling was to find a model for the current dynamics and the polarisation curve which gives the relation between the current and the voltage. The scope of this research is only to find a model of the electrolyser during operation, so start-up and shut-down are interesting, but outside the scope of this paper. 

It is important to note that the model developed here is a closed-loop model of the Enapter module, including its internal control. For system integrators the control of each component from different manufacturers are usually "hidden" and protected by intellectual property rights, so this is a realistic scenario when implementing control of an industrial plant. However, the behaviour of the electrolyser does not necessarily reflect the physical limitations of the electrolyser, as the manufacturer must carefully design a controller of their components that balances multiple aspects such as efficiency, degradation and performance. The electrolyser model developed here does not necessarily reflect dynamics of Enapter electrolysers in the future as they continue to develop and make changes to their products. 

% A model of the electrolyser was deduced based on the data analysis. "we did it like this, and this and this paper did it similar"
\begin{table}
\caption{Parameters for Electrolyser Model}
\label{tab:params}
\begin{center}
\begin{tabular}{|c|c|c|}
\hline
\textbf{Model} & \textbf{Parameter} & \textbf{Value}\\
\hline
\multirow{2}{*}{\textbf{\textit{Polarisation curve}}} & Voltage intercept & 35.75 V\\
\cline{2-3}
& Ohmic resistance & 0.1529 \si{\ohm} \\
\cline{2-3}
& RMSE & 0.4253 V \\
\hline
\multirow{5}{*}{\textbf{\textit{Current dynamic}}} & $K$ & 0.534 s\\
\cline{2-3}
& $\tau_z$ & 95.371 s\\
\cline{2-3}
& $\tau_{p,1}$ & 1.995 s\\
\cline{2-3}
& $\tau_{p,2}$ & 113.568 s\\
\cline{2-3}
& Positive ramp rate limit & 0.22 A/s\\
\cline{2-3}
& RMSE & 0.172 A \\
\hline
\multirow{3}{*}{\textbf{\textit{Hydrogen production}}} & Production coefficient & 9.6174 Nl/hA \\
\cline{2-3}
& RMSE & 0.1297 Nl/h \\
\cline{2-3}
& Nominal current & 53.35 A \\
\hline
\end{tabular}
\end{center}
\vspace{-10pt}
\end{table}

\subsection{Polarisation Curve}
The method to obtain the polarisation curve characteristics is based on the procedure described in \cite{ysteinUlleberg2003ModelingApproach}. The polarisation curve of the electrolyser depends on temperature and pressure, as can be seen in Fig. \ref{fig:dips}. Since these variables are regulated by the manufacturer’s control system, they remain close to their nominal values during operation. Consequently, the available data do not allow a reliable characterization of their influence. The resulting polarisation curve is therefore valid only for the nominal operating conditions (55°C and 29 bar) with the manufacturer’s control system active, implicitly capturing this control behaviour in the model. The area of the membranes are not known.

%%% LISA READ UNTIL HERE%%%

\begin{comment}
Fig. \ref{fig:dips} shows a time series of the temperature of the electrolyte downstream of the stack and the voltage. The dips in temperature are due to the water tank being refilled with colder water from the central water tank. The temperature only deviates from nominal by up to three degrees, making it difficult to obtain a good temperature dependence model. Corresponding spikes in the voltage and small dips in the current can be observed, confirming the temperature dependence of the polarisation curve. 
\end{comment}

The lower operation limit of the electrolyser is at 60\% of nominal hydrogen production, so the logarithmic behaviour of the polarisation curve at low current cannot be observed in the operational data. Hence the polarisation curve is simply modelled linearly. The data points are filtered on temperature and pressure with an absolute tolerance of 0.1 bar and 0.1 °C. 
%Note that module 342A was run the most, which is reflected in the amount of data points. Module A568 was run the least, and thus has less data points so even though the rmse is low it doesn't mean that the fitted line reflects the true polarisation curve.

\begin{comment}
\begin{figure}
\centerline{\includegraphics[width=0.5\textwidth]{Images/combined_linear_IV_TP_windows_scientific_zoomout_short.png}}
\caption{\label{fig:pol}Fitted polarisation curve of each of the AEM modules with only data points from nominal conditions, with an absolute tolerance of 0.1 for both pressure and temperature.}
\vspace{-9pt}
\end{figure}
\end{comment}

\begin{figure}
\centerline{\includegraphics[width=0.45\textwidth]{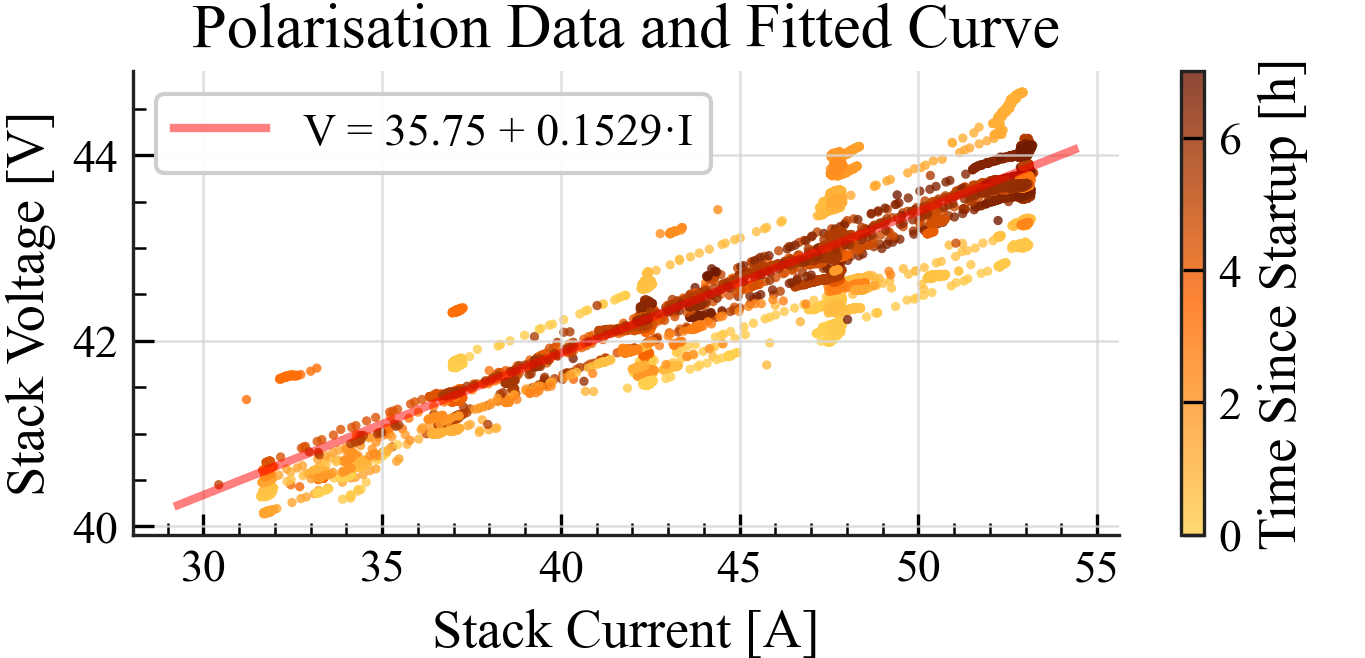}}
\caption{\label{fig:cycle} All polarisation data points of one module during October 2025 filtered on temperature and pressure plotted with shade corresponding to consecutive runtime.}
\vspace{-18pt}
\end{figure}

The RMSE of the data points of the polarisation curve is very large. A steady-state detection algorithm was run to remove transient data points without improving the residual variance. Sorting the data based on electrolyte KOH concentration did also not explain the spread. The best explanation for the spread of data points was found to be related to how long the electrolyser has been running consecutively, as can be seen in Fig. \ref{fig:cycle}. The longer the electrolyser is running, the more the polarisation curve drifts towards a specific behaviour or ohmic resistance.

The polarisation curve identification procedure can be summarised in the following steps:
\begin{enumerate}
    \item Filter current and voltage data points by temperature and voltage
    \item Fit the data points to obtain the polarisation curve
\end{enumerate}
\vspace{-1pt}

\subsection{Current Dynamics}
All the changes in hydrogen production setpoint and the responses were extracted from the data set and sorted into groups based on step size. For the negative step changes, a model order sweep with the subspace identification methods N4SID, MOESP and CVA using the Python Package SIPPY-Unipi was conducted to confirm a model structure candidate \cite{sippy}. The overall best fit was a second order model described by the following transfer function,

\begin{equation}
    G(s) = \frac{K(\tau_z s + 1)}{(\tau_{p,1} s + 1)(\tau_{p,2} s + 1)}
\end{equation}

where $K$ is the gain, $\tau_{p,1}$ and $\tau_{p,2}$ are two pole time constants, and $\tau_z$ is a zero. The positive step responses can be described by this transfer function with an added ramp rate limit.

\begin{figure}
\centerline{\includegraphics[width=0.5\textwidth]{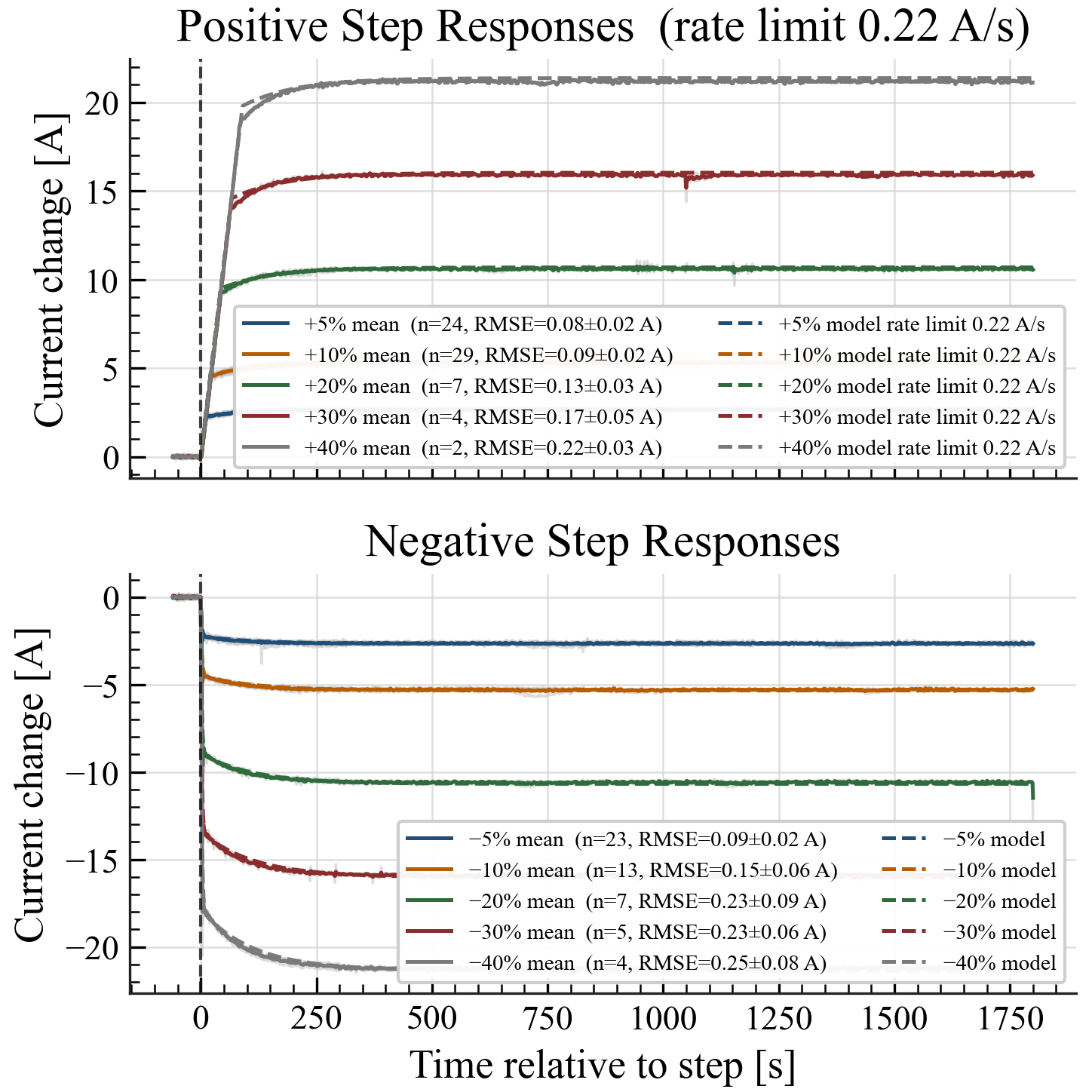}}
\caption{\label{fig:allsteps} Model and validation data of current step responses to positive and negative setpoint changes of varying sizes from 5\% to 40\%. "n" refers to how many steps there is in the validation set of that step size.}
\vspace{-18pt}
\end{figure}

Experimental data from 10 days were used to find the model parameters with uncertainties using a parameter fitting algorithm that minimizes the RMSE across all step sizes. 
The step responses of the dataset were identified and sorted into groups of the same step size. The mean of the current before the step change is subtracted such that all steps starts at zero. Then a parameter fitting algorithm was run to minimize the total RMSE across all step sizes. This was done separately with data from the three electrolyser modules, and separately for positive and negative setpoint changes. 
The validation data and the identified model is shown in Fig. \ref{fig:allsteps}.

%CVA, MOESP and N4SID were tested with a model order sweep on the global data set in a similar way, obtaining one model for positive and one for negative steps using AICc to choose the best model for comparison across models. The RMSE of the transfer function and rate limit model was 0.1 for both positive and negative steps, whereas for the other models the best one was 0.18 for negative steps and 0.4 for positive. There may be more sophisticated ways to tune the algorithms to obtain better models, but that is not the focus of this research. The purpose of these data based models are to make sure the identified model is good enough in comparison. 
\begin{comment}
The RMSE of each model can be found in Tab. \ref{tab:modelrmse}.

\begin{table}
\caption{RMSE of Identified Models}
\label{tab:modelrmse}
\begin{center}
\begin{tabular}{|c||c|c|c|}
\hline
Model & Order & Step direction & Training RMSE [A] \\
\hline
This model & 2 & Positive & 0.1020 \\
\hline
This model & 2 & Negative  & 0.1090\\
\hline
CVA & 3 & Positive & 0.4021  \\
\hline
CVA & 5 & Negative & 0.1779  \\
\hline
MOESP & 4 & Positive & 0.4047  \\
\hline
MOESP & 5 & Negative & 0.1786  \\
\hline
N4SID & 3 & Positive &  0.4053 \\
\hline
N4SID & 5 & Negative & 0.1786  \\
\hline
\end{tabular}
\end{center}
\end{table}
\end{comment}

The identified step response is similar to that based on the experimental data of the alkaline and PEM technologies presented in \cite{NRELAlkPEM} and translated into a transfer function in \cite{MancarellaTransfer}. However, the time constants are quite different, partly due to the individual manufacturers preferred control settings.

The current dynamic model identification procedure can be summarised in the following steps:
\begin{enumerate}
    \item Separate the training data into defined step sizes
    \item Remove outliers
    \item Use step responses to identify the model structure
    \item Apply parameter fitting algorithm to find parameters
    \item Use validation data to benchmark the model
\end{enumerate}

%\subsection{Results from Statistical Analysis}
% Include mean, RMSE etc. 

\subsection{Hydrogen Production Coefficient}
The hydrogen production is proportional to the current, by Faradays Law. Enapters measurement of the hydrogen flow was used to find the relation between current and hydrogen production in Table \ref{tab:params}. 

\subsection{Model Validation}
\begin{figure}
\centerline{\includegraphics[width=0.5\textwidth]{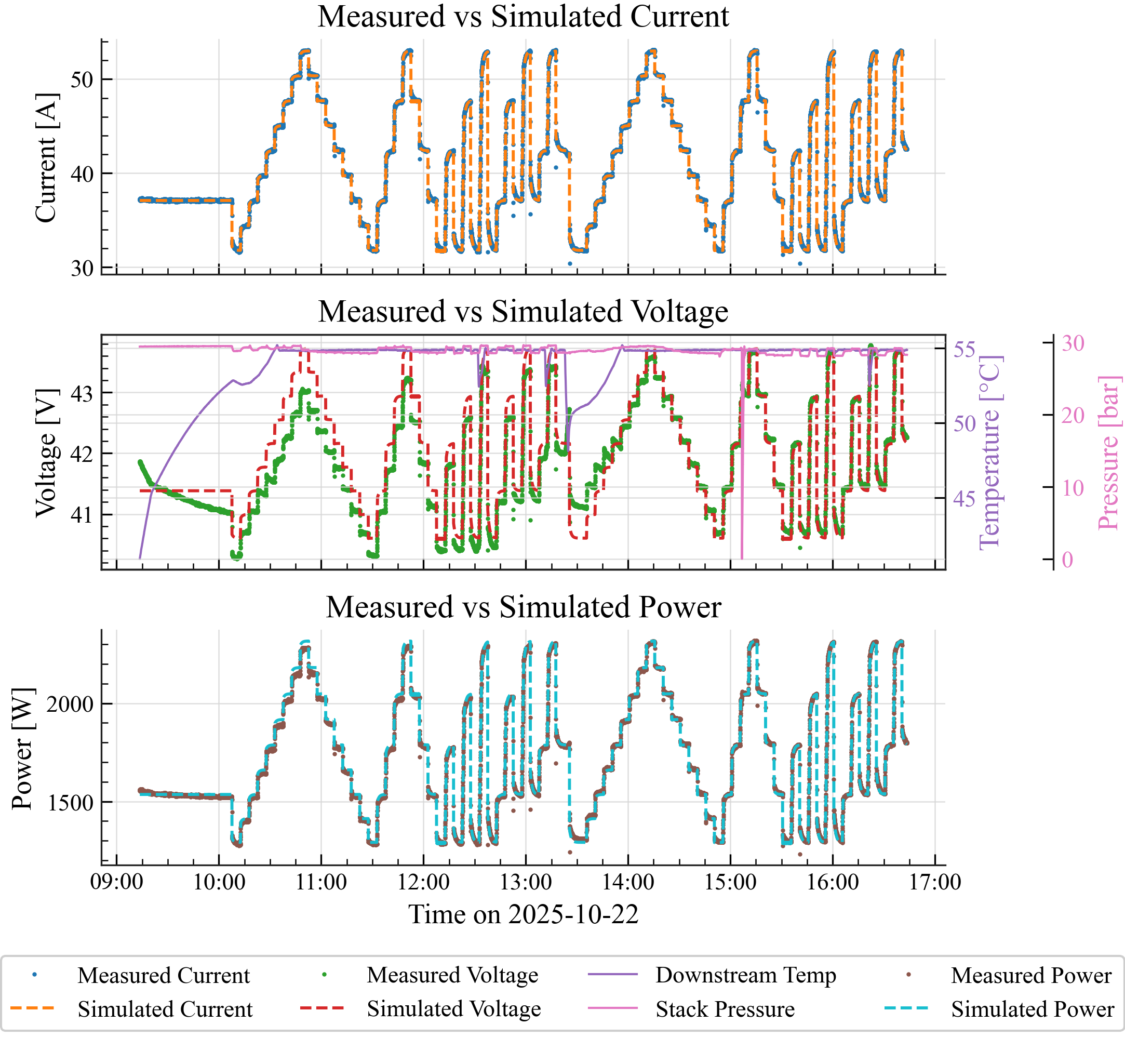}}
\caption{\label{fig:comparison} Validation of model using experimental data from October 22nd. }
\vspace{-10pt}
\end{figure}

The model was validated using experimental data from a given day that was not in the training set and compared with the measured data from that day as shown in Fig. \ref{fig:comparison}. The current model matches well, however the simplified polarisation curve model causes an error in the voltage model in particular right after startup when the electrolyte temperature is still rising. This only marginally affects the simulated DC power, though. Since the current variations are much larger than the voltage variations the effect of a mismatch in voltage is small. Notice that the voltage decreases as the temperature increases until around 10:30, and that the voltage increases when there is a sudden pressure drop at around 15:05. This is not captured by the polarisation curve model presented here, but could be explored in further research.

\vspace{-1pt}

\subsection{Simple Example of Model Use}
The current response and the polarisation curve together capture well the dynamics of the electrolyser during nominal operating conditions. How to implement the model, however, depends on the use case one is interested in, which are probably many and varied. For a dynamic model of the hydrogen production and response characteristics, the transfer function with positive ramp rate limit can be used directly. Using the model in power system simulations, however, requires the modelling of the interface to the grid. 

To demonstrate the use of the model, a green hydrogen production system was implemented in an open source power system simulator in Python, TOPS \cite{TOPS}. This very simple example system includes a wind farm of 64 MW rated capacity, an electrolyser hub rated 60 MW (25000 Enapter modules) and a grid connection. The electrolyser is connected to the grid through a voltage source converter (VSC), where the polarisation curve and current dynamics are translated into an active power reference for the VSC. %Converter losses are neglected.

\begin{figure}
\centerline{\includegraphics[width=0.46\textwidth]{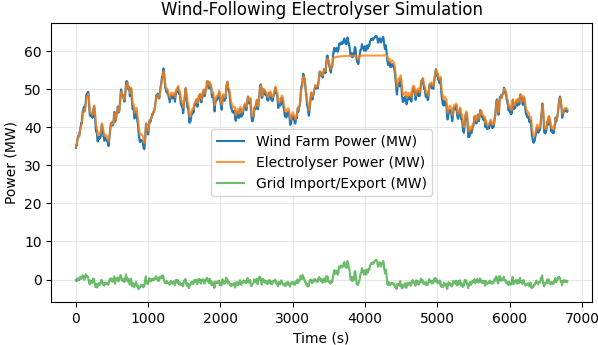}}
\caption{\label{fig:windsim} Simulation of electrolyser modules in wind-following mode.}
\vspace{-18pt}
\end{figure}

Enapters large scale electrolyser containers consist of individually controlled modules, and the same principle is used in this simulation. The wind farm power output is distributed as power references equally to the electrolyser modules in the system. The wind farm power output is modelled as in \cite{cleanoff}, using southern North Sea turbulent conditions with updated values every second. 
Fig. \ref{fig:windsim} shows how one electrolyser module tracks the power reference from the wind farm and its contribution to the grid exchange. Grid exchange is simply calculated as the discrepancy between the wind farm power output and the electrolyser power consumption. 

\begin{comment}
\begin{figure}
\centering
\begin{subfigure}[b]{0.5\textwidth}
  \includegraphics[width=1\linewidth]{Images/electrolyser_current_comparison.png}
  \caption{}
  \label{fig:icomp} 
\end{subfigure}
\medskip % insert a bit of vertical whitespace
\begin{subfigure}[b]{0.5\textwidth}
  \includegraphics[width=1\linewidth]{Images/electrolyser_voltage_comparison.png}
  \caption{}
  \label{fig:vcomp}
\end{subfigure}
\caption[polarisation curve before and after steady state filtration]{\label{fig:comparison}
(a) Current response to varying step sizes of positive changes of setpoint.
(b) Voltage response to varying step sizes of negative changes of setpoint.}
\end{figure}
\end{comment}
\vspace{-1pt}

%%%%%%%%%%%%%%%%%%%%%%%%%%%%%%%%%%%%%%%%%%%%%%%%%%%%%%%%%%%%%%%%%%%%%%%%%%%%%%%%%%%%%%%%%%
\section{Discussion}\label{sec:discussion}
While temperature and pressure factors were examined, additional factors such as outside temperature, degradation, consecutive runtime and operation history may further improve the polarisation curve accuracy. Startup and shutdown behaviours were excluded from the operational model. Incorporating these dynamics would enable application to additional scenarios such as rapid on-off cycling. This would be valuable for large scale facilities composed of hundreds of modules that are controlled to extend the system's lower operational limit by shutting off individual modules.

The rating of the electrolysers in the test setup are rated at 2.4 kW each, and scaling it to utility scale may require different control strategies that are not addressed here. %Hence, it is uncertain how the model will scale with a larger electrolyser capacity. 
Enapter offers containerised solutions where there is a common balance of plant including one common electrolyte tank, pump, heat exchanger and converter %The modules are mounted in stacks of ten, where each one is controlled separately expanding the operational envelope of the container. According to manufacturer specifications, containerised configurations enable operation at minimum loads as low as 1\% compared to 60\% for individual modules 
\cite{EnapterContainer2026}. 
\vspace{-4pt}

%%%%%%%%%%%%%%%%%%%%%%%%%%%%%%%%%%%%%%%%%%%%%%%%%%%%%%%%%%%%%%%%%%%%%%%%%%%%%%%%%%%%%%%%%%
\section{Conclusion}\label{sec:conclusion}
Experimental data from one month of experiments was used to develop a dynamic closed loop model of the Enapter AEM 4.1 stack. The model is composed of a polarisation curve and current dynamics which was identified as a second order transfer function with a ramp rate limit on only the positive steps. The model was validated with satisfactory results for current and power, however the voltage is not accurately modelled due to dynamics that were not captured in the polarisation curve. These dynamics seem to be primarily related to the concecutive runtime of the electrolyser, but also to the temperature and pressure of the stack. 

Further research should include obtaining a polarisation curve model that includes temperature and pressure, and investigate factors such as degradation and runtime, as was discovered had an impact on the polarisation curve every cycle. The model presented here also lacks cold and warm start procedures, which can be essential when controlling a large scale system. The response of the electrolyser should be checked against grid codes and specific requirements of reserve markets. Optimal power control for large scale systems should be investigated, and finally, a model of the power consumption of the auxiliary equipment could be useful to take the whole industrial system into account.

%%%%%%%%%%%%%%%%%%%%%%%%%%%%%%%%%%%%%%%%%%%%%%%%%%%%%%%%%%%%%%%%%%%%%%%%%%%%%%%%%%%%%%%%%%
\vspace{-4pt}
\section{Acknowledgment}
The authors thank PLDK technician Magnus Fich Rabjerg for technical support during the experimental work, and Trond Haugen for valuable discussions.
\vspace{-4pt}

\bibliographystyle{elsarticle-num}

\bibliography{cas-refs}

\end{document}